\documentclass{amsart}
\usepackage{graphicx,bm,color}
\usepackage{setspace}
\usepackage{amssymb,latexsym, amsmath}
\usepackage{verbatim} 
\newcommand*{\ra}{{\rightarrow}}
\newcommand*{\epskm}{{\veps^\alpha_m}}
\newcommand*{\veps}{{\varepsilon}}

\newcommand*{\etakm}{{\eta^\alpha_m}}

\newcommand*{\red}[1]{\textcolor{red}{#1}}

\newcommand*{\ang}[1]{\left\langle #1 \right\rangle}

\newcommand*{\paren}[1]{\left( #1 \right)}
\newcommand*{\muline}{{{\underline{m}}}}
\newcommand*{\moline}{{{\overline{m}}}}

\def\Om{\Omega}
\def\aa{\alpha}
\def\k{k}
\def\ee{\mathfrak{e}}
\def\toprule{\\[-5.5pt]\hline\\[-3.5pt]}
\def\colrule{\\[-7.5pt]\hline\\[-5pt]}
\def\botrule{\\[-7.7pt]\hline}

\begin{document}

\Large{A study of the Navier-Stokes-$\alpha$ model for two-dimensional turbulence}\\
\begin{center}
{\small E. Lunasin$^{A1}$, S. Kurien$^{A2}$, M. A. Taylor$^{A3}$
  and E. S. Titi$^{A1,A4, A5}$}\\
\end{center}
{\small 
$^{A1}$Department of Mathematics, University of California, Irvine, CA, 92697, USA\\
$^{A2}$Theoretical Division, Los Alamos National
  Laboratory, Los Alamos, NM, 87544, USA\\
$^{A3}$Department of Exploratory Simulation Technologies, Sandia
  National Laboratories, Albuquerque, NM, 87185, USA\\
$^{A4}$Department of Computer Science and
  Applied Mathematics, Weizmann Institute of Science, Rehovot, 76100, Israel\\
$^{A5}$Department of Mechanical and Aerospace Engineering, University of California, Irvine, CA, 92697, USA\\
} 

{\small {\sc ABSTRACT.}
The Navier-Stokes-$\alpha$ sub-grid scale model of turbulence is a
  mollification of the Navier-Stokes equations in which the vorticity
  is advected and stretched by a smoothed velocity field.  The
  smoothing is performed by filtering the velocity field over spatial
  scales of size smaller than $\alpha$.  This is achieved by convolution
  with a kernel associated with the Green's function of the Helmholtz
  operator scaled by a parameter $\alpha$.  The statistical properties
  of the smoothed velocity field are expected to match those of
  Navier-Stokes turbulence for scales larger than $\alpha$, thus
  providing a more computable model for those scales. For wavenumbers
  $k$ such that $k\alpha > 1$, corresponding to spatial scales smaller
  than $\alpha$, there are three candidate power laws for the energy
  spectrum, corresponding to three possible dynamical eddy turnover
  time scales in the model equations: one from the smoothed field, the
  second from the rough field and the third from a special combination
  of the two. Using two-dimensional turbulence as a test case, we
  measure the scaling of the spectra from high-resolution simulations
  of the Navier-Stokes-$\alpha$ model, in the limit as $\alpha\ra\infty$. We show that the energy spectrum of the
  smoothed velocity field scales as $k^{-7}$ in the direct enstrophy
  cascade regime, consistent with dynamics dominated by the time scale
  of the {\it rough} velocity field. This result implies that the
  rough velocity field plays a crucial role in the development of the
  smoothed field even though the latter is the purported model for the
  Navier-Stokes turbulence.  This effect must be taken into account
  when performing accurate simulations of the NS-$\alpha$ model.
}
\footnote{This is a preprint of an article submitted for consideration in the Journal of Turbulence, 2007, Taylor \& Francis. }
\section{Introduction\label{intro}}
 Let $v(x,t)$ denote the velocity field and $p(x,t)$ the pressure of a
 homogeneous incompressible fluid with constant density $\rho(x)=1$.
 The Navier-Stokes equations (NSE):
\begin{equation}\label{NSE}
\aligned
\partial_t v   -v\times (\nabla\times v) &= -\nabla p +\nu \Delta v+ f,\\
\nabla \cdot v &= 0,\\
v(x,0) &= v^{in}(x),
\endaligned
\end{equation}
govern the dynamics of such fluid flows, where the body force $f(x)$,
the viscosity $\nu~>~0$, and the initial velocity field $v^{in}(x)$ are given.  We supplement the system in
(\ref{NSE}) with periodic boundary conditions in a basic box
$[0,L]^n$, where $n=2,\mbox{ or } 3$.  We assume here,
$\int_{[0,L]^n} f(x)\; dx = \int_{[0,L]^n} v^{in}(x)\; dx=0$, which implies that
$\int_{[0,L]^n} v(x,t)\; dx=0$ for all $t>0$.
It is
computationally prohibitively expensive to perform reliable direct
numerical simulation (DNS) of the NSE for high Reynolds number flows,
due to the wide range of scales of motion that need to be resolved.
The use of numerical models allows researchers to simulate turbulent
flows of interest using smaller computational resources.  In this
paper we study a particular sub-grid scale turbulence model known as
the Navier-Stokes-$\alpha$ (NS-$\alpha$) model:
\begin{eqnarray}
\aligned
\partial_t v  -u\times\paren{\nabla\times v} &= -\nabla p +\nu\Delta v+ f\\
\nabla\cdot u =\nabla \cdot v&=0\\
v &=u-\alpha^2\Delta u.
\endaligned
\label{2dnsa}
\end{eqnarray}
The system in (\ref{2dnsa}) reduces to the Navier-Stokes system when
$\alpha =0$, i.e. $u = v$.  One can think of the parameter $\alpha$ as
the length scale associated with the width of the filter which smooths
$v$ to obtain $u$.  The filter is associated with the Green's function
of the Helmholtz operator $(I-\aa^2\Delta)$ (given in the third
equation of (\ref{2dnsa})).  When $\alpha>0$, notice that the
nonlinear term $u\times(\nabla \times v)$ in (\ref{2dnsa}) is milder
than that in (\ref{NSE}).

The inviscid and unforced version of the $\alpha$-model in
(\ref{2dnsa}) was derived in \cite{HMR} based on the Hamilton
variational principle subject to the incompressibility constraint
$\mbox{div}\; v =0$. It is worth mentioning that the inviscid model,
the so called Euler-$\alpha$ model, coincides with the inviscid
second-grade non-Newtonian fluid model (see, e.g., \cite{CG,CG2,DF}).
By adding the viscous term $-\nu\Delta v$ and the forcing $f$ in an ad
hoc fashion, the authors in \cite{CH98,CH99,CH00} and \cite{FHTM}
obtain the NS-$\alpha$ system (\ref{2dnsa}) which they then named the
viscous Camassa-Holm equations (VCHE), also known as the Lagrangian
averaged Navier-Stokes-$\alpha$ model (LANS-$\alpha$).  The
LANS-$\alpha$ model is different from the viscous second-grade
non-Newtonian visco-elastic fluid in that the viscous term in the
former is $-\nu\Delta v$, distinguishable from $-\nu\Delta u$ in the
latter.  In references~\cite{CH98, CH99, CH00} it was found that
the analytical steady state solutions for the 3d NS-$\alpha$ model
compared well with averaged experimental data from turbulent flows in
channels and pipes for wide range of Reynolds numbers. It was this
fact which led the authors of \cite{CH98, CH99, CH00} to suggest that
the NS-$\alpha$ model be used as a closure model for the Reynolds
averaged equations (RANS).  It has since been discovered that there is
in fact a whole family of `$\alpha$'- models which provide similar
successful comparison with empirical data -- among these are the
Clark-$\alpha$ model \cite{CHTi, Clark}, the Leray-$\alpha$ model
\cite{CHOT}, the modified Leray-$\alpha$ model \cite{ILT} and the
simplified Bardina model \cite{CLT, LL} (see also \cite{OlTi} for a
family of similar models).

We assume periodic boundary conditions with basic box $[0,L]^n$, and
denote the spatial Fourier coefficients of $u(x)$ and $v(x)$ by $\hat
u(\xi)$ and $\hat v(\xi)$, respectively. Let $(\cdot ,\cdot)$ denote the
$L^2$-inner product and $|\cdot|$ denote the $L^2$-norm over the basic
box $[0,L]^n$ (we also use $|\cdot|$ for the modulus of a vector). In the two-dimensional (2d) case we have two conserved
quantities for the (inviscid and unforced) NS-$\alpha$ equations,
namely the energy,
\begin{eqnarray}\label{ea}
E_\alpha &=& \sum_{|\xi|=1}^\infty E_\alpha(\xi) \mbox{\quad where, }\\
E_\alpha(\xi) &=&
\dfrac{1}{2}\left(\hat{v}(\xi)\cdot\hat{u}(\xi)\right)=\dfrac{1}{2}\left(|\hat{u}(\xi)|^2
+ \aa^2|\xi|^2|\hat{u}(\xi)|^2 \right)\; \;\;
\end{eqnarray}
and the enstrophy,
\begin{eqnarray}\label{oa}
\Omega_\alpha &=& \sum_{|\xi|=1}^\infty \Omega_\alpha(\xi)\\ \mbox{where
}\Omega_\alpha(\xi) &=& \frac{1}{2}|\hat q(\xi)|^2\\ \mbox{~and the
vorticity }\hat q(\xi) &=& (\widehat{\nabla \times v})(\xi).
\end{eqnarray}
The energy of the smoothed field $u$ is given by,
\begin{eqnarray}
E^u &=& \sum_{|\xi|=1}^{\infty} E^u(\xi)\\
\mbox{where }  E^u(\xi) &=& \frac{1}{2}|\hat u(\xi)|^2,
\end{eqnarray}
and is not conserved in the inviscid and unforced equations.  

For a wavenumber $\k$, we define the component $u_\k$ of a velocity field $u$ by
\begin{equation}
u_k:=u_\k (x)= \sum_{|\xi|=\k}  \hat{u}(\xi)e^{i\frac{2\pi}{L}\xi\cdot x},
\end{equation}
and the component $u_{\k',\k''}$ with a range of wavenumbers $[\k',\k'')$ by
\begin{equation}
u_{\k',\k''}:=u_{\k',\k''}(x) = \sum_{\k'\leq\k<\k''}u_\k.
\end{equation}
Denote by $E_\alpha(\k)$ the energy spectrum associated to
\begin{equation}
\mathfrak{e}_\k^\aa = \dfrac{1}{2} \ang{|u_{\k,2\k}|^2+\aa^2|\nabla u_{\k,2\k}|^2},
\end{equation}
which is the average energy per unit mass of eddies of linear size $l\in\biggl(\dfrac{1}{2\k},\dfrac{1}{\k}\biggr]$, then
\begin{equation}\label{ea-spec}
\ee_\k^\alpha =\int_\k^{2\k} E_\alpha(\chi)\;d\chi.
\end{equation}
Similarly, denote by $E^u(\k)$ the energy spectrum associated to $\ee_\k^u =\dfrac{1}{2}\ang{|u_{\k,2\k}|^2}$, that is, 
\begin{equation}
\ee_\k^u = \int_\k^{2\k}E^u(\chi)\;d\chi.
\end{equation} 
In \cite{FHTP}, it can be shown that $E^u(\k)$ has the following remarkable
property in three-dimensional (3d) viscous NS-$\aa$ case,

\begin{eqnarray}
  E^u(\k) \sim \left\{\begin{array}{lll} \k^{-5/3}, \quad &\mbox{when }
                 &\k\alpha \ll 1\\ \k^{-\beta}, \quad &\mbox{when
                 }&\k\alpha \gg 1 \end{array}\right.
\label{beta}
\end{eqnarray}
where $\beta = 3 > 5/3$, corresponds to a sharper roll-off than the
NSE scaling of the energy power spectrum $\k^{-5/3}$ for $\k\alpha\gg
1$.  Thus, $E^u(\k)$ is offered as a model spectrum for the NSE for
spatial scales of motion larger than $\alpha$.  From the point of view
of numerical simulation, the faster (compared to NSE) decay of
$E^u(\k)$ of NS-$\alpha$ for $\k\alpha > 1$, indicates suppression of
scales smaller than $\alpha$ and in principle implies reduced
resolution requirements for simulating a given flow. The scaling in the
$\k\alpha > 1$ regime, that is, the exponent $\beta$, was determined in
\cite{FHTM} using a time scale which depends solely on the smoothed
velocity field.  Later on, in \cite{CHTi,CHOT,ILT}, three
possibilities for the exponent $\beta$ were derived since there are two velocities in the NS-$\alpha$ model.  The three
possibilities for the exponent $\beta$ depend on whether the turnover
time scale of an eddy of size $\k^{-1}$ is determined by
$(\k |u_\k|)^{-1}$, $(\k |v_\k|)^{-1}$, or
$(\k\sqrt{( u_\k, v_\k)}\;)^{-1}$ (see
e.g. \cite{CHTi,ILT,CLT,FHTP} and Section 2).  Determining the actual
scaling requires resolved numerical simulations. In the rest of the
paper we will use the notation $E(\k)$ and $E^u(\k)$ interchangeably for
finite $\alpha$ simulations when there is no ambiguity in meaning.

The 3d NS-$\alpha$ model was tested numerically in \cite{CH01}, for
moderate Reynolds number in a simulation of size $256^3$, with
periodic boundary conditions. It was observed that the large scale
features of a turbulent flow were indeed captured and there was a
rollover of the energy spectrum from $\k^{-5/3}$ for $\k\alpha \ll 1$ to
something steeper for $\k\alpha \gg 1$, although the scaling ranges
were insufficient to enable extraction of the scaling exponent $\beta$
of (\ref{beta}) unambiguously. Other numerical
tests of the NS-$\alpha$ model were performed in
\cite{GH}, \cite{GH2}, and \cite{MM}, with similar results.

Our goal in this work is to measure the scaling of the energy spectrum
of the NS-$\alpha$ in the regime $\k\alpha \gg 1$ in the forward
enstrophy cascade regime for 2d turbulence. Assuming the validity
of the semi-rigorous arguments introduced in
\cite{FHTP}, we can then infer from our computations the timescale
which governs the turnover of an eddy of size $1/\k$. We choose to
do this measurement for the 2d case with the expectation that
discerning scaling ranges and exponents will be cheaper and more
tenable in a 2d system than in a 3d system at the same
grid-resolution. Nevertheless, we hope that the results in 2d will provide
some insight into the turnover time of eddies of size
$1/\k$ for $\k\alpha\gg 1$ in the 3d case.

In two-dimensional turbulence there are two inertial ranges, one
for the inverse cascade of energy $E$ and the other for the direct
cascade of enstrophy  with corresponding
scalings of the energy spectrum as follows \cite{K67}:
\begin{eqnarray*}
E(\k) \sim \Big\{
  \begin{array}{lll} 
    \k^{-5/3},&\quad \mbox{where }&\k \ll \k_f,~\mbox{inverse energy cascade regime}\\
    \k^{-3},&\quad \mbox{where }& \k_d\gg \k \gg \k_f,~\mbox{direct
    enstrophy cascade regime}
  \end{array}
\label{E2d} 
\end{eqnarray*}
where $\k_f$ is the injection (forcing) wavenumber for both energy and
enstrophy, and $\k_d$ is the enstrophy dissipation wavenumber.  The
$\k^{-3}$ scaling is thought to be accurate upto a log-correction
\cite{Klog}. In a NS-$\alpha$ model implementation, if we choose
$\alpha$ smaller than the injection scale such that $\k_f\alpha \ll 1$,
we expect a modification of the scaling in the enstrophy cascade
regime as follows
\begin{eqnarray}
E^u(\k) \sim \left\{\begin{array}{lll} \k^{-5/3}, \quad& \mbox{where }&\k \ll
    \k_f,\\
    \k^{-3}, \quad &\mbox{where }& \k_f \ll \k \ll 1/\aa,  \\
    \k^{-\gamma}, \quad& \mbox{where }&1/\aa \ll \k \ll \k_d \end{array}\right\}
\label{Eu_alpha}
\end{eqnarray}
where $\gamma > 3$.  For the
2d NS-$\alpha$, we show, using semi-rigorous arguments as in \cite{CHTi, FOIAS, ILT, CLT, FHTP} 
that there are three possible scalings, $\gamma \in \{7, 19/3,
17/3\}$. We verify numerically which of these scalings actually emerges, and
therefore deduce the eddy turnover time for the eddies of size smaller than the length scale $\alpha$.

As may be seen from Eqs.~(\ref{Eu_alpha}), for finite $\alpha$, there
are three distinct scaling ranges for 2d NS-$\alpha$.  Since we are
primarily interested in the value of $\gamma$ we would like to
maximize the range of scales over which $\k^{-\gamma}$ scaling
dominates. Therefore, we minimize the inverse cascade range of
$\k^{-5/3}$ scaling by forcing in the lowest modes. We also minimize
the $\k^{-3}$ scaling range by increasing $\alpha$ all the way to
$\alpha \rightarrow \infty$. The latter limit is well-defined and
yields what we will call the NS-$\infty$ equations (see,
e.g. \cite{IT}): 
\begin{eqnarray}\label{2dns-inf}
\aligned
\partial_t v  -u\times\nabla\times v &=
-\nabla p +\nu\Delta v+f\\
\nabla\cdot u =\nabla\cdot v&=0\\
v &= -L^2\Delta u,
\endaligned
\end{eqnarray}
where the forcing term was rescaled appropriately to avoid trivial
dynamics.  Assuming that the scaling of the spectrum as $\alpha
\rightarrow \infty$ is identical to the scaling in the range $\k\alpha
\gg 1$, for finite (small) $\alpha$ and sufficiently long scaling
ranges, we obtain the first resolved numerical calculation of the
sub-$\alpha$ scales. The assumption is not unreasonable as nothing
singular is expected to occur as $\alpha$ grows, other than the
lengthening of the scaling range of interest.  We remark, however,
that one has to rescale the forcing appropriately in order to avoid
trivial dynamics for large values of $\aa$, and decaying turbulence at
the limit when $\aa\ra\infty$.

The paper is organized as follows. In section \ref{math} we summarize
our analytical results. We show that the 2d NS-$\alpha$ model exhibits
an inverse cascade of the energy $E_\alpha = \frac{1}{2}\int_{[0,L]^2}
u(x)\cdot v(x)\; dx$ and a forward cascade of the enstrophy
$\Omega_\alpha = \frac{1}{2}\int_{[0,L]^2} |q(x)|^2\; dx$, where
$q(x)~=~\nabla~\times~ v(x)$.  Then, we show that in the forward
enstrophy cascade regime, the energy of the smoothed velocity $E^u =
\frac{1}{2}\int_{[0,L]^2} u(x)\cdot u(x)\; dx$ should follow the
Kraichnan $\k^{-3}$ power law for $\k\alpha \ll 1$.  For $\k$ such
that $\k\alpha \gg 1$ and $\k_f \ll \k \ll \k_d$, we will show how the
three possible power laws, $\k^{-17/3}, \k^{-19/3}$ and $
\k^{-7}$, arise.

In section \ref{results} we present the details of our numerical
scheme, data parameters and empirical justification for adopting a
hypoviscousity sink term for the energy in the low modes.  Our 
numerical calculations ranged in resolution from $256^2$ to $4096^2$. 
Our data show a convergence to $\k^{-7}$ power law
in the range $\k\alpha \ll 1 $ as $\alpha \rightarrow \infty$. This
scaling allows us to conclude that the
eddy turnover time in the enstrophy cascade region of the smoothed 2d
NS-$\alpha$ energy spectrum is determined by the time scale which
depends on the square of the unsmoothed velocity field, that is the
variable $v$. 

\section{Statistical predictions  for the NS-$\alpha$ model \label{math}}

\subsection{Average transfer and cascade of energy and enstrophy for the 2d NS-$\alpha$ model}\label{scaling}
In two-dimensional turbulence, energy and enstrophy transfer behave as
follows: in the wavenumber range below $\k_f$, the energy and enstrophy
go from higher modes to lower modes; in the wavenumber range above
$\k_f$, the energy and enstrophy go from low to high wavenumbers.  In a
certain wavenumber regime above $\k_f$, there is much stronger direct
transfer of enstrophy than energy, leading to what is called the direct
enstrophy cascade.  Similarly, in a certain range below $\k_f$, there
is a more dominant transfer of energy, leading to what is called
the inverse energy cascade.  For the 2d NS-$\alpha$ model we show that
there is similar behaviour of transfer and cascade for its energy
$E_\alpha$ and enstrophy $\Omega_\alpha$.  We follow the techniques in
\cite{FMRT} to arrive at this conclusion.  Here we summarize our main
results.  The complete details can be found in the Appendix.

For $m\gg \k_f$, let $\epskm$ be the net amount of energy $E_\alpha$
(see (\ref{ea})) transferred per unit time into the modes higher than
or equal to $m$, and $\etakm$ be the net amount of enstrophy
$\Omega_\alpha$ (see (\ref{oa})) transferred per unit time into the
modes higher than or equal to $m$. From the scale-by-scale evolution
equation of energy $E_\alpha$ and enstrophy $\Omega_\alpha$ we get the
following relationship between $\epskm$ and $\etakm$
\begin{eqnarray}
\ang{\epskm} \leq 
\dfrac{\ang{\etakm}}{m^2
(1+\aa^2m^2)},
\end{eqnarray}
where $\ang{\cdot}$ denotes an ensemble average with respect to infinite time average measure (see Appendix). This result suggests that for large $m$, the
average net of transfer of energy to high modes is significantly
smaller than the corresponding transfer of enstrophy.  This yields the 
characteristic direct enstrophy cascade.

The inverse energy cascade is expected to take place in the range below the
$\k_f$. Within this
range, i.e. $m\ll \k_f$, we obtain
\begin{equation}
\ang{-\etakm}\leq m^2(1+\aa^2m^2)\ang{-\epskm},
\end{equation}
that is, the (inverse) average net transfer of energy to lower modes
is much stronger than the corresponding enstrophy transfer which
yields the characteristic inverse energy cascade. This establishes the
expected 2d NS-$\alpha$ cascades for energy $E_\alpha$ and enstrophy
$\Omega_\alpha$.
 
\subsection{NS-$\alpha$ model effects on the scaling of the energy spectrum
  in the enstrophy cascade regime\label{scaling}}
We next derive, using semi-rigorous arguments \cite{FOIAS,FHTM}, the
expected scaling for the 2d NS-$\alpha$ energy spectrum. We start
by splitting the flow into the three
wavenumber ranges $[1,\k), [\k,2\k),[2\k,\infty)$. We assume $\k_f < \k$,
where $\k_f$ is the forcing wavenumber, since we are interested on the
effects of the NS-$\alpha$ model in the enstrophy cascade regime. We
decompose the $u$, $v$ and vorticity $q$  
corresponding to the three wavenumber range, and for simplicity, we denote as follows:
\begin{eqnarray*}
u = u_\k^< + u_{\k,2k} + u_{2\k}^>\\
v = v_\k^< + v_{\k,2k} + v_{2\k}^>\\
q = q_\k^< + q_{\k,2k} + q_{2\k}^>,
\end{eqnarray*}
where,
$\phi_l^< = \phi_{1,l}$, $\phi_l^> = \phi_{l,\infty}$.  We recall $\paren{\cdot,\cdot}$ and $|\cdot|$ denote the
$L^2$-inner product and $L^2$-norm respectively.  The enstrophy
balance equation for the NS-$\alpha$ model for an eddy of size
$\sim \k^{-1}$ is given by
\begin{eqnarray}
&&\hskip-.28in
\frac{1}{2}\; \frac{d}{dt} (q_{\k,2k}, q_{\k,2k}) +\nu (-\Delta q_{\k,2k}, q_{\k,2k}) = Z_\k -Z_{2\k},
\label{BAL}
\end{eqnarray}
where,
\begin{eqnarray*}
\aligned
Z_\k &:= -b(u_\k^<,q_\k^<,q_{\k,2k}+q_{2\k}^>) + b(u_{\k,2k} + u_{2\k}^>, q_{\k,2k} + q_{2\k}^>, q_\k^<)\\
-Z_{2\k} &:= -b(u_{2\k}^>, q_{2\k}^>, q_{\k}^<+q_{\k,2k})+b(u_\k^<+u_{\k,2k}, q_\k^< + q_{\k,2k},
q_{2\k}^>)\\
b(u,v,w) &:=
\paren{u\cdot\nabla v, w}.
\endaligned
\end{eqnarray*}
$Z_\k$ may be interpreted as the net amount of enstrophy per unit time that is
transferred into wavenumbers larger than or equal to $\k$. Similarly, $Z_{2\k}$
represents the net amount of enstrophy per unit time that is transferred
into wavenumbers larger than or equal to $2\k$. Thus, $Z_\k -
Z_{2\k}$ represents the net amount of enstrophy per unit time that is
transferred into wavenumbers in the interval $[\k,2\k)$.  Taking an ensemble
average (with respect to infinite time average measure) of (\ref{BAL}) we get 
\begin{equation}\label{TAET1}
\nu \ang{(-\Delta q_{\k,2k},q_{\k,2k})} = \ang{Z_\k} -\ang{Z_{2\k}}. 
\end{equation}

Using the definition of $E_\alpha(\k)$ in (\ref{ea-spec}), we can
rewrite the averaged enstrophy transfer equation (\ref{TAET1}) as
$$\nu \k^5(1+\alpha^2\k^2)E_\alpha(\k) \sim \nu \int_\k^{2\k} \k^4(1+\alpha^2\k^2)E_\alpha(\k)d\k \sim
\ang{Z_\k}-\ang{Z_{2k}}.$$ 
Thus, as long as $\nu \k^5(1+\alpha^2\k^2)E_\alpha(\k) \ll \ang{Z_\k}$ (that is,
$\ang{Z_{2\k}}\approx\ang{Z_\k}$, there is no leakage of enstrophy due to
dissipation), the wavenumber $\k$ belongs to the inertial range.
Similar to the other $\alpha$-models, the correct averaged velocity of
an eddy of length size $\sim 1/k$ is not known.
That is, we do not know {\it a priori} in these models the exact eddy
turnover time of an eddy of size $\sim 1/k$.  

In the forward cascade inertial subrange, Kraichnan theory \cite{K67} postulates that the eddies of size larger than $1/k$ transfer their energy to eddies of size smaller than $1/(2k)$ in the time $\tau_k$ it takes to travel their length $\sim 1/k$.  That is,
\begin{equation}
\tau_\k \sim \frac{1}{\k U_\k},
\label{tau_}
\end{equation}
where $U_\k$ is the average velocity of eddies of size $\sim 1/\k$. Since
there are two velocities in the NS-$\alpha$ model, there are three physically relevant possibilities for this average velocity, namely
\begin{eqnarray*}\label{3-vel}
  &&\hskip-.28in
  U_\k^{0}
  = \ang{ \frac{1}{L^2} \int_{\Om} |v_{\k,2k}(x)|^2
    dx}^{1/2} \sim \left(\int_\k^{2\k}(1+\alpha^2\k^2)E_\alpha(\k)dk\right)^{1/2}\sim \left( \k (1+\alpha^2 \k^2) E_{\alpha} (\k) \right)^{1/2},   \\
  &&\hskip-.28in
  U_\k^{1} = \ang{ \frac{1}{L^2} \int_{\Om} u_{\k,2k}(x) \cdot v_{\k,2k}(x)
    dx}^{1/2} \sim \left(\int_\k^{2\k}E_\alpha(\k)d\k\right)^{1/2}\sim \left( \k E_{\alpha} (\k) \right)^{1/2},   \\
  &&\hskip-.28in
  U_\k^{2} = \ang{ \frac{1}{L^2} \int_{\Om} |u_{\k,2k}(x)|^2
    dx}^{1/2} \sim \left(\int_\k^{2\k}\frac{E_\alpha(\k)}{(1+\alpha^2\k^2)}d\k\right)^{1/2} \sim \left( \frac{\k E_{\alpha} (\k) }{ 1+\alpha^2 \k^2}
  \right)^{1/2}.
\end{eqnarray*} 
That is,
\begin{equation}\label{n}
U_\k^n \sim \frac{(\k E_\alpha(\k))^{1/2}}{(1+\alpha^2\k^2)^{(n-1)/2}}, ~~ (n
= 0,1,2).
\end{equation}
We may therefore write the turnover time scale of an eddy of size $\k^{-1}$ in (\ref{tau_}) as 
\begin{equation}
\tau_\k^n \sim \frac{1}{\k U_\k^n}
 = \frac{(1+\alpha^2\k^2)^{(n-1)/2}}{\k^{3/2}(E_\alpha(\k))^{1/2}}, ~~(n = 0, 1, 2).
\label{taun}
\end{equation}

The enstrophy dissipation rate $\eta_\alpha$ which is a constant
equal to the flux of enstrophy from wavenumber $\k$ to $2\k$ is given by 
\begin{equation}
\eta_\alpha \sim \frac{1}{\tau_\k^{n}} \int_k^{2\k} (1+\alpha^2\k^2)\k^2 E_{\alpha} (\k) d\k \sim
\frac{\k^{9/2} \left( E_{\alpha} (\k) \right)^{3/2} }{(1+\alpha^2
\k^2)^{(n-3)/2}},
\end{equation}
and hence
\[
E_{\alpha}(\k) \sim \frac{\eta_\alpha^{2/3} (1+\alpha^2\k^2)^{(n-3)/3}}{\k^{3}}.
\]
Thus, the energy spectrum of the smoothed velocity $u$ is given by 
\begin{eqnarray}\label{3-spec}
&&\hskip-.28in
E^u(\k)
\equiv
\frac{E_{\alpha} (\k)}{1+ \alpha^2 \k^2} \sim \left\{
\begin{array}{ll}   \displaystyle{
{\eta_{\alpha}^{2/3}}{\k^{-3}},}
\qquad & \mbox{when  }
\k\alpha \ll 1\,, \\
\displaystyle{\frac {\eta_{\alpha}^{2/3}}{\alpha^{2(6-n)/3}}
\k^{-(21-2n)/3}}, \qquad & \mbox{when  } \k\alpha \gg 1\,.
\end{array} \right.
\end{eqnarray}
Therefore, depending on the average velocity of an eddy of size
$\k^{-1}$ for the NS-$\alpha$ model, we obtain three possible scalings
of the energy spectrum, $\k^{-(21-2n)/3}$, $(n = 0,1,2)$ all of which
decay faster than the Kraichnan $\k^{-3}$ power law, in the
subrange $\k\alpha \gg 1$. We note that in the work of \cite{NS} 
the $\k^{-17/3}$ power law was obtained  using
dimensional analysis,
consistent with $n = 2$
in our notation, using a time scale which depends solely on the
smoothed velocity field. The actual power law was not
determined at the time due to insufficient dynamic range of their
simulations. Our simulations here show that the
scaling is $\k^{-7}$, which corresponds to $n=0$ in our notation.

\section{Results\label{results}}
\subsection{Details of the numerical simulation\label{num}}
The Navier-Stokes equations and the Navier-Stokes-$\alpha$ model
equations with stochastic forcing were solved numerically in a
periodic domain of length $L=1$ on each side. The wavenumbers $\k$ are
thus integer multiples of $2\pi$. A pseudospectral code was
used with fourth-order Runge-Kutta time-integration. Simulations were
carried out with resolutions ranging from $256^2$ to a maximum
resolution of $4096^2$ on the Advanced Scientific Computing QSC machine
at the Los Alamos National Laboratory.  To maximize the enstrophy
inertial subrange, the forcing is applied in the wavenumber shells $2
< \k < 4$. We also add a hypoviscous term ($\mu\Delta^{-1}v$) which
provides a sink in the low wavenumbers. Let $\mathcal{H}= 1-\aa^2\Delta$, then $v =\mathcal{H}u$.  
We simulate the following equation:
\begin{eqnarray}\label{tau-form}
\partial_t u - \mathcal{H}^{-1}(u\times (\nabla\times v)) =- \nabla \pi + \mu(\Delta^{-1}u)+ \nu\Delta u + f
\end{eqnarray}
where $\pi$ denotes the modified pressure and $\mu > 0$ is a hypoviscosity coefficient.  The equation in (\ref{tau-form}) is equivalent to  
\begin{eqnarray}
\partial_t v -u\times\paren{\nabla\times v} &= -\nabla p + \mu(\Delta^{-1}v) + \nu\Delta v + \tilde{f},
\label{2dnsa-hypo}
\end{eqnarray}
where $\tilde{f} = \mathcal{H}(f)$ (compare with equation (\ref{2dnsa})). 
We demonstrate in the next
section, that adding a hypoviscous term does not have a significant
effect on our observed numerical results.  We use no extra dissipation
term in the small scales besides the regular dissipation in the
Navier-Stokes or the Navier-Stokes-$\alpha$.  Some relevant parameters
and results of our simulation are presented in Table \ref{symbols}.
\begin{table}[ht]
  \caption{Parameters of the simulations: number of grid points per side
  N, $\alpha$ in units of $\Delta x$, viscocity coefficient $\nu$,
  hypoviscosity coefficient $\mu$, maximum wavenumber $\k_{max}= \dfrac{\sqrt{2}}{3}N$
  , enstrophy dissipation scale $l_\eta = \left(\dfrac{\nu^3}{\eta_\alpha}\right)^{1/6}$, where $\eta_\alpha$ is the enstrophy dissipation rate.}
\centering
  {\begin{tabular}{@{}lccccc}\toprule N & $\alpha$ & $\alpha/L$
  &$\nu$&
  $\mu$ & $\k_{max}l_\eta$ \\
\colrule
 256 &0 &0   &$1e^{-3}$  & 0 &10   \\
     && & & 5 &10   \\
     &&    & & 10&10\\
    &2.04& 0.008     & & 0 &10 \\
 &    & && 5 & 10  \\
 &    & & &10 & 10  \\
 & $\infty$& $\infty$    & &0 &   2\\
&   & & &5 & 2 \\
&    & && 10 & 2  \\\\
1024 & 0& 0 & $1e^{-4}$&15& 12.5\\
& 3.25&0.003 &&&12.5\\
&15&0.015 &&&12\\
&100& 0.097&&&6\\
&$\infty$&$\infty$&&&2\\\\
2048&0&0&$5e^{-5}$& 10&18\\
&6.5&0.003&&& 18 \\
&$\infty$&$\infty$&&&2\\\\
4096&0&0&$1e^{-5}$&10&15\\
&$\infty$&$\infty$ && & 2\\
\botrule
\end{tabular}}
\label{symbols}
\end{table}
 The $256^2$
simulations were performed to ascertain whether it was feasible to
perform the computations in reasonable time without a low-wavenumber
sink term, the hypoviscosity. The simulations of $1024^2$ and higher
show the convergence of the scaling exponent of interest, namely the
value of $\gamma$.

The condition for a resolved simulation is judged by the degree to
which the enstrophy dissipation scale is resolved since the enstrophy
cascade is expected to govern the dynamics in the range $\k \gg \k_f$. The
enstrophy dissipation rate for NSE and NS-$\alpha$ are:
\begin{eqnarray}
  \aligned
  \eta&= \eta_\alpha = \nu\langle |\nabla q|^2 \rangle
\endaligned
\end{eqnarray}
where $q = \nabla \times v$, and $v$ is respectively, either 
the NS velocity or the unfiltered velocity for the NS-$\alpha$.
The enstrophy dissipation scale is then computed, following Kraichnan 
\cite{K67,Klog}, by dimensional
analysis in a manner analogous to the way in which the 
Kolmogorov energy dissipation scale is calculated in 3d turbulence:
\begin{eqnarray}\label{leta}
l_{\eta_{[\alpha]}} = \left(\dfrac{\nu^3}{\eta_{[\alpha]}}\right)^{1/6}.
\end{eqnarray}
A resolved simulation has $\k_{max}l_\eta > 1$. In all our
simulations, $\k_{max}l_\eta \geq 2$, indicating well-resolved
flow. Note, however, in table \ref{symbols} that, keeping all other
parameters fixed, increasing $\alpha$ decreases $\k_{max}l_\eta$,
indicating that as if the NS-$\alpha$ flow is less resolved, from the
point of view of the enstrophy cascade, than the NSE for a given grid
and viscosity.  However, this observation is misleading since the
computations for the NS-$\alpha$ were done with a rescaled forcing
term with an amplitude growing indefinitely as $\alpha\ra\infty$.
Therefore, by increasing $\alpha$, the NS-$\alpha$ is forced more
vigorously, and damped strongly.  We discuss this further in section
(\ref{disslength}).

\subsection{Dependence of scaling behaviour on
  hypoviscosity\label{hypodep}}
\begin{figure}
\centering
\includegraphics[scale=.6]{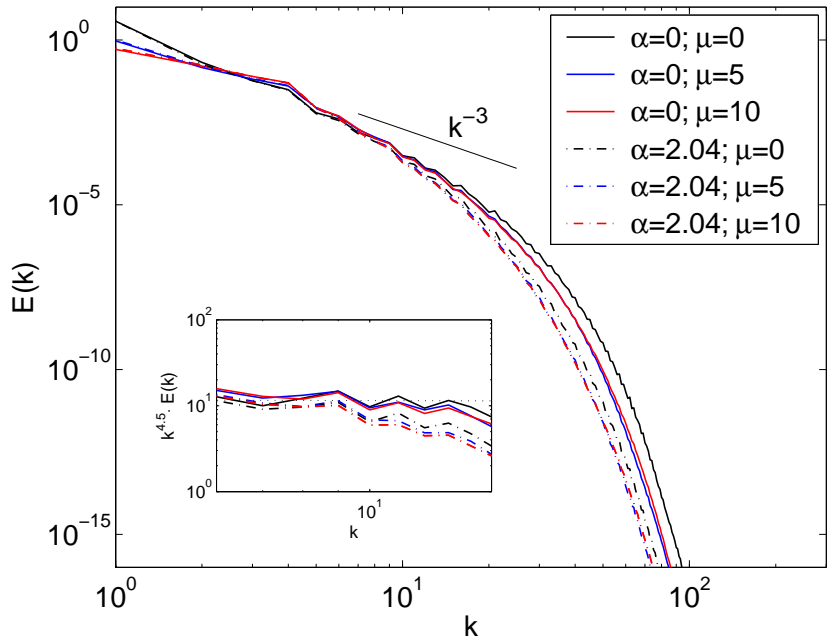}
\caption{Energy spectra for a $256^2$ simulation with fixed
  viscosity and varying hypoviscosity coefficient $\mu$. The wavenumber $\k$
  is in multiples of $2\pi$. The solid lines are
  the DNS ($\alpha=0$) calculations of $E(\k)$. The dotted lines are the NS-$\alpha$ model
  calculations of $E^u(\k)$ for small $\alpha$. The behaviour of the spectra is
  largely independent of the magnitude of the hypoviscosity in the
  enstrophy cascade subrange ($6 < \k < 15$). The inset shows the
  spectra compensated by $\k^{4.5}$. The resolution of this simulation
  is far to small to observe the expected scaling
  exponent. }
\label{figure1}
\end{figure}

The hypoviscosity term in (\ref{2dnsa-hypo}) provides a sink in
the low wavenumbers for the energy. This allows the flow to reach
statistical equilibrium in more reasonable computational time.
However, since it is an ad hoc addition to the NSE or NS-$\alpha$
model, we need to ascertain whether it affects the behaviour in the
range of scales of interest, namely the range $\k> 1/\alpha$.

In figure \ref{figure1} we compare the energy spectra for $256^2$ run
for DNS $(\alpha = 0)$ and 2d NS-$\alpha$ with $\alpha = 2.04$ (in
units of $\Delta x$).  The solid lines represent the DNS runs with
varying hypoviscosity coefficient. In the inset, the compensated plots
for the DNS show very little difference. The dotted lines correspond
to the 2d NS-$\alpha$ runs with varying hypoviscosity coefficient.
Again, as seen in the compensated plots in the inset, there is very
little dependence on the hypoviscosity. These numerical results show
that for a small to moderate hypoviscosity coefficient, the spectral
slope in the enstrophy inertial range is not significantly affected by
the addition of the hypoviscous term.  Furthermore, as we see in
Table \ref{symbols} for the $256^2$ simulation, the enstrophy
dissipation length scale $l_\eta$ is not affected by varying
hypoviscosity coefficient.

From the above empirical observation, we conclude that, for the range of scales
of interest, we can safely use a hypoviscous term and save significant
computational time in our higher resolution runs.

\subsection{Varying alpha; convergence to NS-$\infty$} 

\begin{figure}[ht]
\centering
\includegraphics[scale=.6]{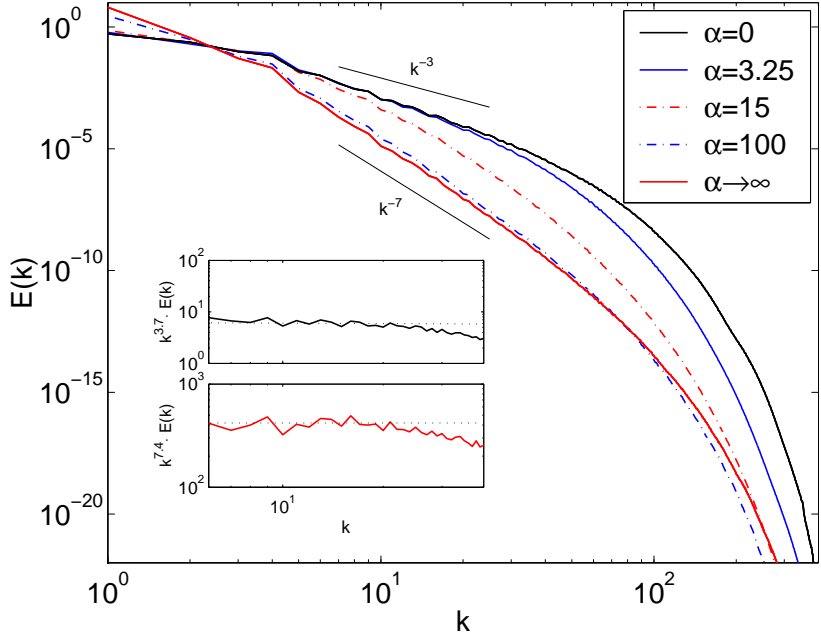}
\caption{Energy spectra for $1024^2$ simulation. The black curve is
 the DNS ($\alpha = 0$) which shows close to $\k^{-3}$ scaling in the enstrophy
  cascade range $6 < \k < 20$. The solid red curve is the $E^u(\k)$ spectrum as
$\alpha \rightarrow \infty$ which scales close to $\k^{-7}$ in the enstrophy
cascade range $6< \k < 25$. The energy spectra
for intermediate values of $\alpha$ tend to the $\alpha
\rightarrow \infty$ limit as $\alpha$ increases. The inset shows the
DNS energy spectrum (black) compensated by $\k^{3.7}$ and the 
$\alpha \rightarrow \infty$ energy spectrum (red) compensated by
$\k^{7.4}$} 
\label{figure2}
\end{figure}

\begin{figure}[ht]
\centering
\includegraphics[scale=.6]{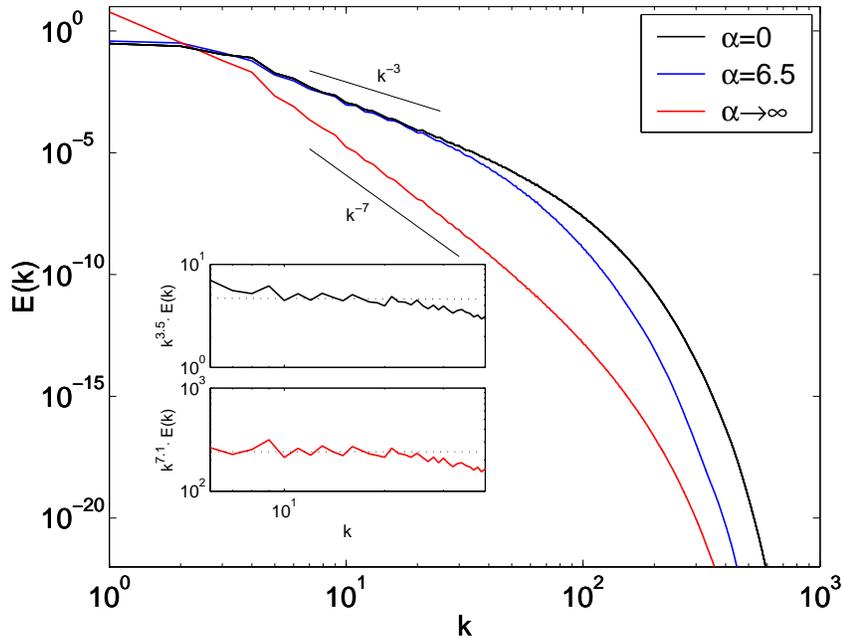}
\caption{Energy spectra for $2048^2$ simulation. The wavenumber is
  in multiples of $2\pi$. The black curve is the energy spectrum of the
  DNS which shows close to $\k^{-3}$ scaling in the enstrophy cascade
  range $6<\k<35$. The solid red curve is the $E^u(\k)$ spectrum as
  $\alpha \rightarrow \infty$ which scales approximately as $\k^{-7.1}$
  in the wavenumber region $6<\k<25$. The inset shows the
DNS energy spectrum (black) compensated by $\k^{3.5}$ and the 
$\alpha \rightarrow \infty$ energy spectrum (red) compensated by
$\k^{7.1}$}
\label{figure5}
\end{figure}

\begin{figure}[ht]
\centering
\includegraphics[scale=.6]{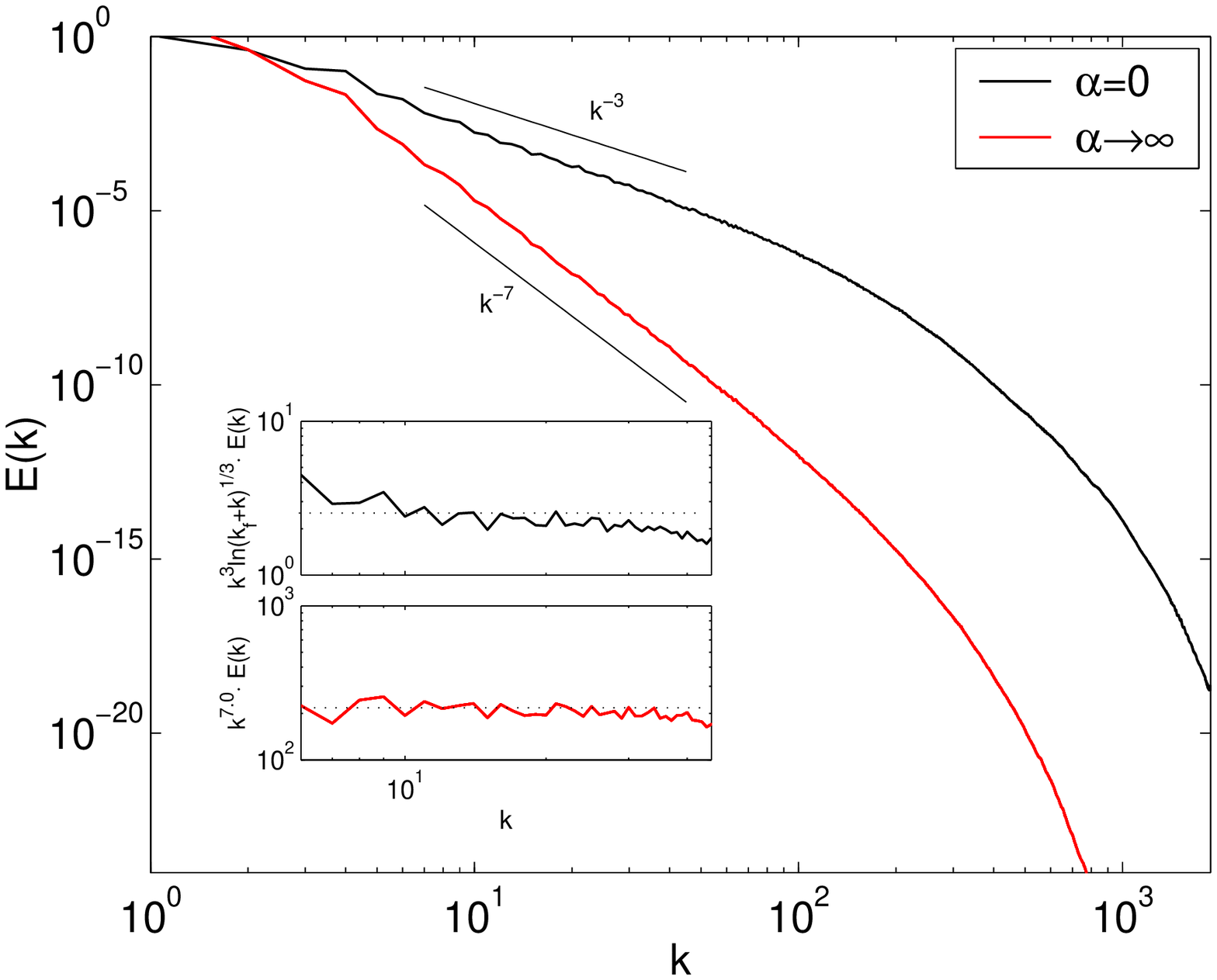}
\caption{Energy spectra for $4096^2$ simulation. The black curve is
  the spectrum for the DNS, the red curve is the spectrum for $\alpha
  \rightarrow \infty$.  The black curve in the inset corresponds to
  the NSE energy spectrum compensated by $\k^3\ln(\k_f+\k)^{1/3}$
  following \cite{Klog}. The red curve in the inset is the energy
  spectrum $E^{u}(\k)$ for NS-$\infty$ compensated by $\k^{7}$. The
  region $6 < \k < 40$ is flat indicating the nominal range over which
  the $\k^{-7}$ scaling holds.  }
\label{4096a_1e20}
\end{figure}

In this section we show the effect of varying the parameter
$\alpha$. For a given grid, we fix the viscosity coefficient and vary
$\alpha$ in order to measure the scaling exponent $\gamma$ of the
energy spectrum in the enstrophy inertial subrange.

In figures \ref{figure2}, \ref{figure5} and \ref{4096a_1e20} we show
the NSE spectrum $E(\k)$ and the spectrum $E^u(\k)$ from $1024^2$,
$2048^2$ and $4096^2$ simulations, each with fixed viscosity and varying
$\alpha$.  As expected, the scaling ranges increase as the
number of grid-points increase. In each figure the solid black line is the DNS
spectrum $E(\k)$, and approaches a scaling close to $\k^{-3}$ as $N$
increases. In both Figs.  \ref{figure2} and \ref{figure5} for $0<
\alpha \leq 15$ (in units of $\Delta x$), we see the spectrum $E^u(\k)$
peels away from the NSE spectrum ($\alpha = 0$) near $\k\alpha = 1$ but
displays no clear scaling behaviour for $\k\alpha > 1$ until $\alpha
\geq 100$. To discern a clear power-law of the NS-$\alpha$ model
spectrum, we consider the data from simulation of the NS-$\infty$
equations (\ref{2dns-inf}). This allows us to see the scaling of the
NS-$\alpha$ model energy spectrum without contamination by
finite-$\alpha$ or DNS for the NSE ($\k^{-3}$) effects.  At our maximum resolution
of $4096^2$ in Fig.~\ref{4096a_1e20}, as $\alpha \ra \infty$ (solid
red line), there is a clear convergence of the scaling to $\k^{-7}$.
Table \ref{infty-table} summarizes our findings for the scaling of the
$E^u(\k)$ spectrum as $\alpha \ra \infty$ for each of our simulations.
\begin{table}[ht]
  \caption{Convergence of $\alpha\ra\infty$ scaling exponent $\gamma$ of
    $E^u(\k)$ as
    the resolution is increased.}
\centering
  {\begin{tabular}{ccccc}\toprule N & 256 & 1024 & 2048&
      4096\\
      $\gamma$ & 8.0 & 7.4 & 7.1& 7.0 \\
      \botrule
\end{tabular}}
\label{infty-table}
\end{table}    
According to (\ref{3-spec}), the scaling $\k^{-7}$ corresponds to an
enstrophy turnover time scale determined by the velocity $v$. We conclude
that the dynamics of the smoothed velocity field, which is the
puported model for turbulence, is nevertheless still governed by the
unfiltered velocity field. 

\begin{figure}[ht]
\centering
\includegraphics[scale=.6]{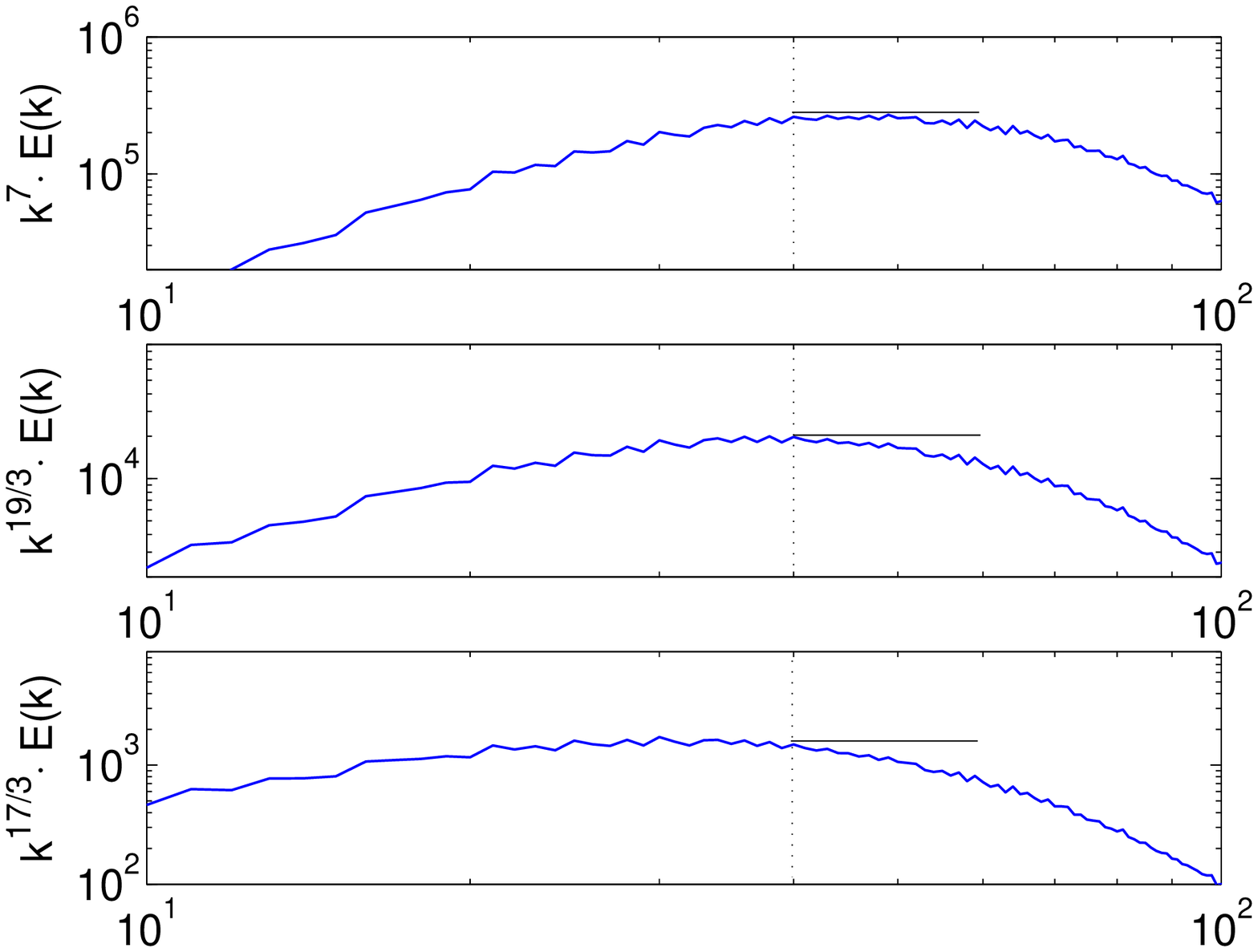}
\caption{Compensated energy spectra for $2048^2$ simulation for
  $\alpha =6.5$ $(\k_\alpha=39.75$; vertical dashed line).  The energy spectrum is
  compensated by $\k^7, \k^{19/3}, \mbox{ and }, \k^{17/3}$ respectively.
  The region $39 < \k < 70$ in the first subplot follows a flat regime
  which indicates the nominal range over which the
  $\k^{-7}$ scaling holds.
 }
\label{2048a6.5}
\end{figure}
In this analysis and conclusion, we have assumed that the scaling in
the asymptotic limit $\alpha \ra \infty$ would be identical to the
scaling for $\k\alpha > 1$ for small $\alpha$. In Fig.~\ref{2048a6.5}
we demonstrate that this assumption is reasonable by showing that the
scaling of the energy spectrum for small $\alpha = 6.5$ is approaching
close to $\k^{-7}$ scaling for a small range of $\k\alpha > 1$. We are
therefore reasonably convinced that going for the asymptotic limit is
indeed giving us the correct scaling for finite (small) $\alpha$; the
$\alpha \ra \infty$ limit merely maximizes the range over which a
clear scaling exponent can be measured at a given resolution.

\subsection{NS-$\alpha$ model effects on the dissipation length scales of
  the flow\label{disslength}} In Table \ref{symbols}, we observe that,
  keeping the viscosity, hypoviscosity and the resolution fixed,
  increasing $\alpha$ tends to decrease the enstrophy dissipation
  length-scale $l_\eta$ (see (\ref{leta})). We also learned that the
  rate of dissipation of the energy $E^{u}$ is dictated by the eddy
  turnovertime of the unfiltered velocity field.  In this section, we
  further explore the effects of the $\alpha$ parameter on the
  smallest scales of the flow.  We recall that the NS-$\alpha$
  computations were done with a rescaled forcing term, a fact which is important in the 
discussion to follow.

\begin{figure}[h]
\centering
\hspace*{-1.5cm}
\includegraphics[scale=.4]{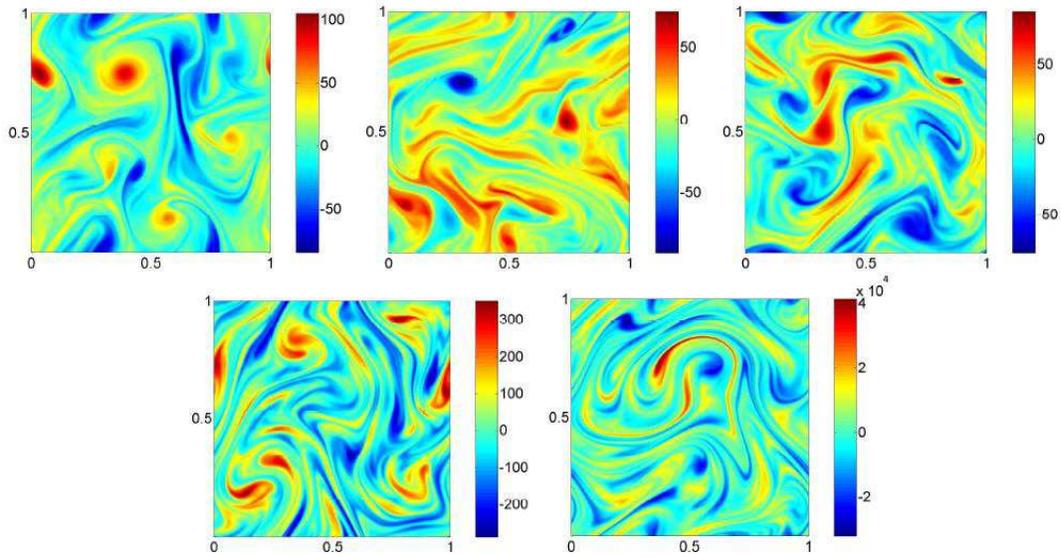}

\caption{Isosurfaces of vorticity $\nabla\times v$ for the $1024^2$
  simulation. $\alpha = 0, 3.25, 15, 100, \infty$,  reading each row
  of figures from left to right. The vorticity field
  exhibits increasingly fine structures as $\alpha$ is increased. }
\label{curlv}
\end{figure}

\begin{table}
  \caption{Dissipation length scales when varying $\alpha$.  N = resolution, $\alpha$, $\nu$ = visocisty coefficient,
  $\mu$ - hypoviscosity coefficient, 
  $l_\eta$- enstrophy ($\Omega_\alpha$) dissipation length scale,
$l_{\eta_{u}}$-smoothed enstrophy dissipation length scale}
\centering
{\begin{tabular}{@{}lcccc|ccccc}\toprule
   N  & $\alpha$ (in units of $\Delta x$) & $\nu$& $\mu$ & $l_\eta$& N  &
   $\alpha^{\rm a}$ & $\nu$& $\mu$ & $l_{\eta_{u}}$ \\
\colrule
 1024 &0    &$1e^{-4}$ & 15  &.004133 &1024 &0  &$1e^{-4}$& 15  &.004133
 \\
      &3.25    & &   &.004099 &&3.25  &&   &.004490   \\
  &15    & &   &.003973 &&15 &&   &.005659   \\
  &100    & &   &.002165 &&100  &&   &.006827   \\
  &$\infty$    & &   &.000488 &&$\infty$  &&   &.006858   \\
 \botrule
\end{tabular}}
\label{dissipation}
\end{table}
In the left panel of table \ref{dissipation}, we present the enstrophy 
dissipation length scale $\l_\eta$ (see (\ref{leta})) for the
$1024^2$ simulation as $\alpha$ increases. This length scale, as we
already know, gets smaller with increasing $\alpha$ value. A visual of
this effect is given in Fig. \ref{curlv}, where we plot the
isosurfaces of vorticity for increasing values of $\alpha$. Observe
how the vorticity values grow while the vorticity structures become
much finer as $\alpha$ increases.

\begin{figure}[h]
\centering
\hspace*{-1.5cm}
\includegraphics[scale=.4]{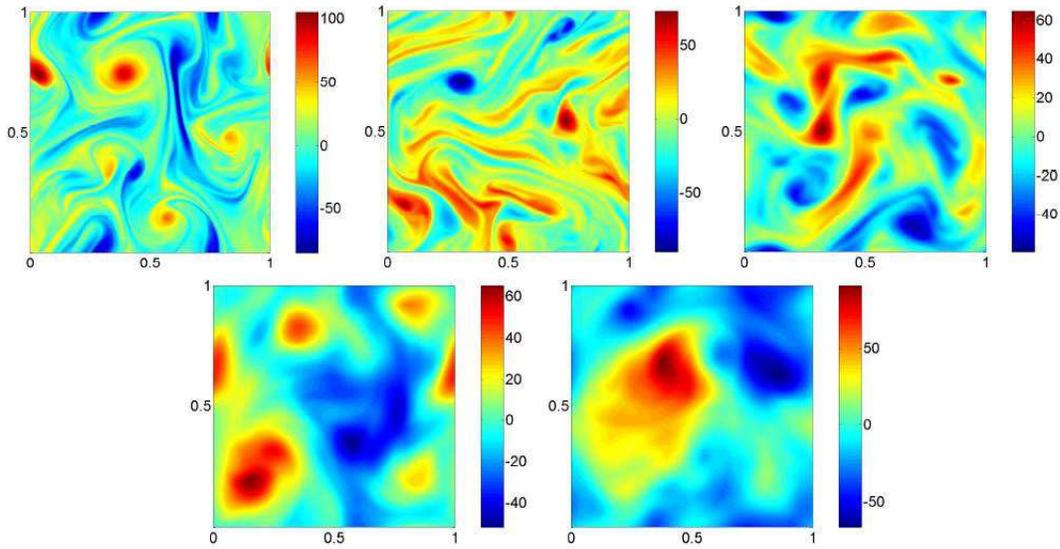}
\caption{Isosurfaces of vorticity $\nabla\times u$ for the $1024^2$
  simulation. 
$\alpha = 0, 3,25,
  15, 100, \infty$, reading each row of figures from left to right.
  The structures become smoother with increasing $\alpha$. }
\label{curlu}
\end{figure}

By contrast, consider the right panel of table \ref{dissipation} where
we present a ``dissipation'' length scale 
\begin{eqnarray}
l_{\eta_u} = \left(\dfrac{\nu^3}{\eta_u}\right)^{1/6},\\
\mbox{ where } \eta_u = \nu\ang{|\nabla\omega|^2},\\
\mbox{ and  } \omega = \nabla\times u.
\end{eqnarray}
corresponding to a naively calculated length-scale for the
smooth enstrophy $\Omega^u = |\nabla \times
u|^2$. This length-scale grows as $\alpha$ increases. Corresponding to
this we present in Fig. \ref{curlu} the isosurfaces of $\nabla \times
u$ for increasing values of $\alpha$. Note that in this case, the
vorticity values diminish while the vorticity structures become
increasingly smooth and diffuse. 

Thus, on the one hand the smooth velocity field and its vorticity are
consistent with {\it reduced} resolution requirements for the
NS-$\alpha$ model; on the other, the behaviour of the {\it conserved}
quantity for the (inviscid and unforced) NS-$\alpha$ model indicate a
requirement for {\it increased} number of grid points, counter to
what one would like to see in a sub-grid model. Note, however, that
in order to avoid trivial dynamics as $\alpha$ increases, we
have scaled the forcing term in the simulations of the
NS-$\alpha$. That is, the computed Eq.~(\ref{2dnsa-hypo}) tends
to the Eq.~(\ref{2dns-inf}) as $\alpha \rightarrow \infty$. This was
done so that we could conveniently study the case of large $\alpha$,
thus extending the scaling range of interest. It could well be that for
a detailed study of small-$\alpha$ with the forcing unscaled, the
desired computational gains expected of a sub-grid model could be
observed. However, our main goal in this study, the scaling exponent
$\gamma$, could not be clearly obtained at achievable resolutions
without going for the large-$\alpha$ limit.  Thus, a straightforward
comparison, for the purpose of checking the NS-$\alpha$ as a sub-grid
model, is not clear. This implies that the implementation of the
NS-$\alpha$ model as a turbulence model needs to be performed with
some care.

\section*{Acknowledgments}
We are grateful to B. Nadiga and B. Wingate for useful discussions and
helpful comments on this study. This work was carried out under the
auspices of the National Nuclear Security Administration of the U.S.
Department of Energy at Los Alamos National Laboratory under Contract
No.  DE-AC52-06NA25396, partially supported by the Laboratory Directed
Research and Development Program and the DOE Office of Science
Advanced Scientific Computing Research (ASCR) Program in Applied
Mathematics Research. This work was also supported in part
by the NSF grant no.~DMS-0504619, the ISF grant no.~120/06, the BSF
grant no.~2004271 and the US Civilian Research and Development
Foundation, grant no.~RUM1-2654-M0-05.

\appendix

\label{appendix}
\section{Average transfer and cascade of the energy $E_\alpha$ and
  enstrophy $\Omega_\alpha$ in the two-dimensional NS-$\alpha$ model}\label{transfer}
Here we will show analytically, the transfer and cascade
of the energy in the two-dimensional
NS-$\alpha$ model.  Following the exposition as in \cite{FMRT}, we will
show that the conserved (in the absence of viscosity and forcing) energy $E_\alpha$ (\ref{ea}) and enstrophy
$\Omega_\alpha$ (\ref{oa}) in the two-dimensional NS-$\alpha$ model have similar
transfer and cascade behaviour as in the two-dimensional turbulence.  We recall that we denote by $\phi_l^< = \phi_{1,l}$ and $\phi_{l}^> = \phi_{l,\infty}$
\begin{enumerate}
\item {\it Average energy transfer}\\
 We start by decomposing the velocity fields into two components:
the large-scale component $u_m^<$ and  $v_m^<$ containing eddies of size larger than $1/m$ and the
small-scale component $u_m^>$ and $v_m^>$ containing eddies of size smaller than or equal to the lengthscale $1/m$.
That is, 
\begin{eqnarray*}\label{decomp}
\aligned
u &= u_m^< + u_m^>\\
v &= v_m^< + v_m^>.\\
\endaligned
\end{eqnarray*}
 For the forcing $f$, we
assume it to be time independent and contains only finite number of
modes, namely, $$f = f_{\muline,\moline},
\mbox{ where, } \; 1< \muline < \moline < \infty.$$
Here we will look at the two cases: $m\in (1,\muline)$ and $m \in (\moline,\infty)$.  
We recall that
$\paren{\cdot,\cdot}$ and $|\cdot|$, denote the $L^2$-inner product and
$L^2$-norm, respectively.
From the above decomposition, we define the kinetic energy 
contained in the large and small eddies (respectively) as 
\begin{eqnarray*}
E^<_\aa = \dfrac{1}{2}\paren{u_m^<,v_m^<}\;\;\mbox{ and }\;\;
E^>_\aa = \dfrac{1}{2}\paren{u_m^>,v_m^>}.
\end{eqnarray*}
We are interested on how the energy $E_\aa$ is transferred between
large and small scales. Consider the case where $m > \moline$
(i.e. beyond the injection of energy).  Following the notation as in
\cite{FHTM},  let us denote by
$\tilde{b}(u,v,w) = \paren{-u\times(\nabla\times v), w}$.  We write (\ref{2dnsa}) in its functional differential form (see,e.g. \cite{FMRT, TT84}),and apply the large and small scale decomposition to get the
corresponding energy equations for the large and small scales
\begin{eqnarray}
\aligned
\dfrac{d}{dt}\paren{v_m^<,u_m^<} + \nu \paren{v_m^<, -\Delta u_m^<} +\tilde{b}(u_m^<+u_m^>,v_m^<+v_m^>,u_m^<)&=\paren{f,u_m^<}\\
\dfrac{d}{dt}\paren{v_m^>,u_m^>} + \nu \paren{v_m^>, -\Delta u_m^>} +\tilde{b}(u_m^<+u_m^>,v_m^<+v_m^>,u_m^>)&=0.
\endaligned
\end{eqnarray}
Define
\begin{eqnarray}
\aligned
\Phi^< &:= -\tilde{b}(u_m^>,v_m^>,u_m^<) +\tilde{b}(u_m^<,v_m^<,u_m^>)\\
\Phi^> &:= \tilde{b}(u_m^>,v_m^>,u_m^<) -\tilde{b}(u_m^<,v_m^<,u_m^>)\\
\endaligned
\end{eqnarray}
Notice that
$\Phi^> + \Phi^< = 0.$
We can rewrite the energy equations in the following form
\begin{eqnarray}\label{energy-phi}
\aligned
\dfrac{d}{dt}E_\aa^< +\nu \paren{v_m^<,-\Delta u_m^<} &= \Phi^< +\paren{f,u_m^<}\\
\dfrac{d}{dt}E_\aa^> +\nu \paren{v_m^>,-\Delta u_m^>} &= \Phi^>.
\endaligned
\end{eqnarray}
We can interpret the individual terms above as follows:
$\paren{f,u_m^<}$ - represents the energy flow injected into the large
scales, by the forcing term.  $\Phi^<$ - represents the net amount
of energy per unit time that is transferred from small to large
length scales.  $\Phi^>$ - represents the net amount of energy per
unit time that is transferred from large to small length scales.
$-\tilde{b}(u_m^<,v_m^<,u_m^>)$ - represents the energy flow induced in
the high modes by inertial forces associated with lower modes. 

Let $\mathcal{F}$ be the set of all vector trigonometric polynomials
with periodic domain $\Omega$.  We then set $\mathcal{V} =
\{\phi\in\mathcal{F}:\nabla\cdot\phi=0 \mbox{ and } \int_\Omega
\phi(x)\;dx=0\}$.  We set $V$ to be the closure of $\mathcal{V}$ in
the Sobolev space $H^{1}$.  The solution operator $u(t) = S_\alpha(t)u_0$ of (\ref{2dnsa}) defines a dynamical system on the phase space $V$.  A generalized notion of {\it infinite time averaging}, as it was defined in \cite{FMRT}, induces a probability invariant measure with respect to $S_\alpha(t)$.  We denote this measure by $dP$, and the ensemble average with respect to this measure, will be denoted by
\begin{equation}\label{dp}
\ang{\phi(\cdot)} = \int_V\phi(w)dP(w).
\end{equation}
Thus, beyond the range
where energy is injected, the average transfer of energy is into the
higher modes, that is 
\begin{eqnarray}\label{eps-forward}
\ang{\Phi^>} = \nu\ang{v_m^>,-\Delta u_m^>}\ge 0.
\end{eqnarray}
Now consider modes that are below the injection of energy.  For that
purpose, assume $\muline > 1$ and consider $m$ such that $1< m < \muline$.
The NS-$\aa$ equation (\ref{2dnsa}) can be decomposed as
follows:
\begin{eqnarray}
\aligned
\dfrac{d}{dt}\paren{v_m^<,u_m^<} + \nu \paren{v_m^<, -\Delta u_m^<} +\tilde{b}(u_m^<+u_m^>,v_m^<+v_m^>,u_m^<)&=0\\
\dfrac{d}{dt}\paren{v_m^>,u_m^>} + \nu \paren{v_m^>, -\Delta u_m^>} +\tilde{b}(u_m^<+u_m^>,v_m^<+v_m^>,u_m^>)&=\paren{f,u_m^>}.
\endaligned
\end{eqnarray}
The associated energy equation for the lower modes reads:

$$\dfrac{d}{dt}E_<^\aa + \nu\paren{v_m^<,-\Delta u_m^<} =\Phi^<.$$
We then take the ensemble average to get
\begin{eqnarray}
\ang{\Phi^<} = \nu\ang{v_m^<,-\Delta u_m^<}\ge 0.
\end{eqnarray}
That is, for wavenumber regime below the injection of energy the average transfer of energy is from high wavenumbers to lower
wavenumbers. \\
\item {\it Average enstrophy transfer}\\
Next we consider the details of the transfer of
enstrophy.  Again, we assume $f$ is time independent and contains only
a finite number of modes.  Here, we do a slightly different exposition
as in \cite{FMRT}.  We take the curl of the momentum equation and using the incompressibility condition and the vector identity,
\begin{equation}
\nabla\times(a\times b) = a(\nabla\cdot b) - b(\nabla\cdot a)-(a\cdot\nabla)b +(b\cdot\nabla)a,
\end{equation}
we get the vorticity formulation for the 2d NS-$\alpha$ equation:
\begin{equation}
\partial_t q -\nu \Delta q + u\cdot\nabla q = \nabla\times f
\end{equation}
where,
$q = \nabla \times v$ and we recall that $v=u-\aa^2\Delta u$. We split the flow into two components
$u = u_m^< + u_m^> \mbox{ and } q = q_m^< + q_m^>$.

The amount of enstrophy contained in the large and small eddies are
given, respectively, by
\begin{eqnarray}
\aligned
\Omega_\aa^< &= |\nabla\times v_m^<|^2 \\
\Omega_\aa^> &= |\nabla\times v_m^>|^2. 
\endaligned
\end{eqnarray} 
We would like to show how the enstrophy is transferred between the
large and small scales.  First consider the case $m> \moline$
(i.e. beyond the injection of energy). Following the notation as in
the NSE \cite{FMRT}, we denote by $b(u,v,w) = \paren{u\cdot\nabla
  v,w}$.  We write the evolution of the
enstrophy  governing the large and small scales:
\begin{eqnarray}
\aligned
\dfrac{d}{dt}\paren{q_m^<,q_m^<} + \nu \paren{q_m^<, -\Delta q_m^<}
+b(u_m^<+u_m^>,q_m^<+q_m^>,q_m^<)&=\paren{\nabla\times f,q_m^<}\\
\dfrac{d}{dt}\paren{q_m^>,q_m^>} + \nu \paren{q_m^>, -\Delta q_m^>} +b(u_m^<+u_m^>,q_m^<+q_m^>,q_m^>)&=0.
\endaligned
\end{eqnarray}
where, we denote by
\begin{eqnarray}
\aligned
\Psi^<&:=-b(u_m^<+u_m^>,q_m^<+q_m^>,q_m^<) = -b(u_m^<,q_m^<,q_m^>)+b(u_m^>,q_m^>,q_m^<)\\
\Psi^>&:=-b(u_m^<+u_m^>,q_m^<+q_m^>,q_m^>) = b(u_m^<,q_m^<,q_m^>)-b(u_m^>,q_m^>,q_m^<) 
\endaligned
\end{eqnarray}
$\Psi^<$ - is the net amount of enstrophy transferred into low
modes and $\Psi^>$ - is the net amount of enstrophy transferred into high
modes.  Note that for almost every $t$, $\Psi^< +\Psi^> = 0$. From this we can rewrite the enstrophy equations
above in the following form
\begin{eqnarray}
\aligned
\dfrac{1}{2}\dfrac{d}{dt}|q_m^<|^2 + \nu|\nabla q_m^<|^2 &= \Psi^< +
\paren{\nabla\times f, q_m^<}\\
\dfrac{1}{2}\dfrac{d}{dt}|q_m^>|^2 + \nu|\nabla q_m^>|^2 &= \Psi^>.
\endaligned
\end{eqnarray}  
Taking the ensemble average (with respect to the infinite time averaging measure $dP$) of the equation for enstrophy associated with
the low modes, we get
\begin{equation}\label{eta-forward}
\ang{\Psi^>} =\nu\ang{|\nabla q_m^>|^2}\ge 0.
\end{equation}
This implies that, beyond the injection of energy, the average transfer of enstrophy is from low modes into higher modes.  Next we consider the modes below the injection
of energy.  Assume $1< m < \muline$, we get
\begin{eqnarray}
\aligned
\dfrac{d}{dt}\paren{q_m^<,q_m^<} + \nu \paren{q_m^<, -\Delta q_m^<}
+b(u_m^<+u_m^>,q_m^<+q_m^>,q_m^<)&=0\\
\dfrac{d}{dt}\paren{q_m^>,q_m^>} + \nu \paren{q_m^>, -\Delta q_m^>}
+b(u_m^<+u_m^>,q_m^<+q_m^>,q_m^>)&=\paren{\nabla\times f,q_m^>}.
\endaligned
\end{eqnarray}
We take the ensemble average of enstrophy equation associated with the
low modes to get
$$ \ang{\Psi^<} = -\ang{\Psi^>} = \nu |\nabla q_m^<|^2\ge 0,$$
that is, the average net transfer of enstrophy is from high modes to lower
modes for wavenumber regime below the injection of energy.\\
\item {\it Direct enstrophy cascade and inverse energy cascade}\\
In (i) and (ii) we have seen that in the range above the injection of
energy both the energy and enstrophy is transferred from low to higher
wavenumbers.  On the wavenumbers regime below
the injection of energy, both the energy and enstrophy are transferred
from high to lower wavenumbers.  Here we will show that, in the wavenumbers regime above the injection of
energy, there is a much
stronger transfer of enstrophy leading to what is called direct
enstrophy cascade . On the other hand, in the wavenumbers regime below the
injection of energy, the energy transfer
is much stronger than the enstrophy transfer leading to what is
known as the inverse cascade of energy.

We split the flow into three parts assuming the same form restriction
on the forcing as above.  Consider $m'' \geq m'\geq \moline$.  Let $u = u_{m'}^< +u_{m',m''} + u_{m''}^>$ and $q = q_{m'}^< +q_{m',m''} + q_{m''}^>$.  The vorticity field associated with the wavenumbers between $m'$ and $m''$ is
\begin{eqnarray*}
\dfrac{d}{dt}q_{m',m''} -\nu \Delta q_{m',m''} + B(u_{m'}^<+u_{m',m''}+u_{m''}^>,q_{m'}^<+q_{m',m''}+q_{m''}^>) = 0
\end{eqnarray*}
where we denote by $B(v,w):= (v\cdot\nabla) w$.
The evolution equation for the enstrophy associated with the modes
between $m'$ and $m''$ is given by
\begin{eqnarray*}
\aligned
\dfrac{1}{2}\dfrac{d}{dt}|q_{m',m''}|^2 &+ \nu|\nabla q_{m',m''}|^2 \\&=  -b(u_{m'}^<+u_{m',m''}+u_{m''}^>,q_{m'}^<+q_{m',m''}+q_{m''}^>, q_{m',m''} ),
\endaligned
\end{eqnarray*}
where, $b(u,v,w) := ((u\cdot\nabla) v,w)$.
One can compute by using the bilinear properties of $b$ (see, e.g. \cite{CF88, TT84, FMRT}) that, 
\begin{eqnarray}
\aligned
-b(u_{m'}^<+u_{m',m''}+u_{m''}^>,q_{m'}^<+q_{m',m''}+q_{m''}^>, q_\k ) &= \eta_{m'}^\aa - \eta_{m''}^\aa
\endaligned
\end{eqnarray}
where,\\ 
$\eta^\aa_{m'}$ - is the net amount of enstrophy transferred per unit time into the modes higher than or equal to $m'$ and $\eta^\aa_{m''}$ - is the net amount of enstrophy transferred per unit time into the modes higher than or equal to $m''$, and given by
\begin{eqnarray*}
\aligned
\eta^\aa_{m'}&= -b(u_{m'}^<, q_{m'}^<, q_{m',m''}+q_{m''}^>) + b(u_{m',m''}+u_{m''}^>, q_{m',m''}+q_{m''}^>,q_{m'}^<) \\
-\eta^\aa_{m''}&= -b(u_{m''}^>, q_{m''}^>, q_{m'}^<+q_{m',m''}) + b(u_{m'}^<+u_{m',m''}, q_{m'}^<+q_{m',m''},q_{m'}^>). 
\endaligned
\end{eqnarray*} 
In explicit form we have
\begin{eqnarray}\label{ens-k}
\dfrac{1}{2}\dfrac{d}{dt}|q_{m',m''}|^2 +\nu|\nabla q_{m',m''} |^2 = \eta^\aa_{m'} - \eta^\aa_{m''}.
\end{eqnarray}
Take the ensemble average of (\ref{ens-k}), we get
 $$\ang{\eta^\aa_{m''}}=\ang{\eta^\aa_{m'}}-\nu\ang{|\nabla q_{m',m''}|^2}.$$
That is, the average net transfer of enstrophy into the modes higher
than $m''$ is equal to the average net transfer of enstrophy into
the modes higher than or equal to $m'$ minus the enstrophy lost due
to viscous dissipation within the range $[m',m'']$.  In
$[m',m'']$, it is assumed that the viscous dissipation is negligible and the
enstrophy is simply transferred to smaller and smaller eddies.  This occurs whenever
\begin{eqnarray}\label{inertial}
 \ang{\eta^\aa_{m''}}\ge\ang{\eta^\aa_{m'}}\gg\nu\ang{|\nabla q_{m',m''}|^2}
\end{eqnarray}
The range of wavenumbers $\k>\moline$ up to where (\ref{inertial})
holds is called the enstrophy inertial subrange.  Now within this range there is still an average net transfer of
energy to higher modes.  Denote by $\ang{\epskm} := \ang{\Phi^>}$ and
$\ang{\etakm} := \ang{\Psi^>}$.  From (\ref{eps-forward}) and (\ref{eta-forward})
\begin{eqnarray*}
\dfrac{1}{\nu}\ang{\epskm} \leq \dfrac{\ang{|q_{m,\infty}|^2}}{1+\aa^2m^2}\leq  \dfrac{\ang{|\nabla q_{m,\infty}|^2}}{m^2(1+\aa^2m^2)} = \dfrac{1}{\nu}\dfrac{\ang{\etakm}}{m^2(1+\aa^2m^2)}
\end{eqnarray*}
Hence, 
\begin{equation}
\ang{\epskm} \leq \dfrac{\ang{\etakm}}{m^2(1+\aa^2m^2)}.
\end{equation}
This result suggests that for large $m$, in particular, $\k_f<m<\k_d$, where $k_d$ is the dissipation wavenumber, the average net of
transfer of energy to high modes is significantly smaller than the
corresponding transfer of enstrophy.  This yields the characteristic
direct enstrophy cascade in this range.

The inverse energy cascade takes place in the range below the injection of
energy.  Consider $1< m'< m''< \muline$.  We
follow the same steps as above.  We decompose the flow into three components $u = u_{m'}^< + u_{m',m''} +u_{m''}^>$ and proceeding as before we obtain
\begin{equation}
\ang{\veps^\aa_{m''}} =
\ang{\veps^\aa_{m'}}+\nu\ang{\paren{v_{m'm''},-\Delta u_{m',m''}}}
\end{equation}
As we have seen in (i) we have in this case the inverse transfer of energy
$$ \ang{\veps^\aa_{m''}}\leq 0\; \mbox{ and }\; \ang{\veps^\aa_{m'}}\leq 0 $$
Therefore, as long as
\begin{equation}
\ang{\veps^\aa_{m''}} \lesssim
\ang{\veps^\aa_{m'}}\ll\nu\ang{v_{m',m''},-\Delta u_{m', m''}}
\end{equation}
we have inverse cascade.  Now within the range corresponding to energy
cascade (below injection of the force $f$) both the energy and enstrophy are transferred to lower modes
\begin{eqnarray*}
\aligned
\ang{-\epskm} &= \nu\ang{\paren{\nabla v_{1,m}, \nabla
    u_{1,m}}} = \nu\ang{|\nabla u_{1,m} |^2 + \aa^2|\Delta  u_{1,m}|^2}\ge 0\\
\ang{-\etakm}&=\nu\ang{|\nabla q_{1,m}|^2}\ge 0
\endaligned
\end{eqnarray*}
Since $u_{1,m}$ contains only modes smaller than $m$, we
therefore have
\begin{equation}
|\nabla q_{1,m}|^2\leq m^2(1+\aa^2m^2)\paren{\nabla v_{1,m}, \nabla u_{1,m}}
\end{equation}
which implies
\begin{equation}
\ang{-\etakm}\leq m^2(1+\aa^2m^2)\ang{-\epskm}
\end{equation}
that is, in the wavenumber regime below the injection of energy, the (inverse) average net transfer of energy to lower modes is
much stronger than the corresponding enstrophy transfer which yields
the characteristic inverse energy cascade.
\end{enumerate}

\end{document}
